\documentclass[twocolumn,aps,showpacs,amsmath,reprint,amssymb,superscriptaddress,prl]{revtex4-1}

\usepackage[FIGTOPCAP,tight]{subfigure} %FIGTOPCAP
\usepackage{amsmath, amssymb}
\usepackage[T1]{fontenc}
\usepackage[latin9]{inputenc}
\usepackage{geometry}
\usepackage{graphicx}
\usepackage{esint}
\usepackage{array}
\usepackage{multirow}

\begin{document}

\title{Revealing Hidden Coherence in Partially Coherent Light}

\author{Ji\v{r}\'{i} Svozil\'{i}k}
\email{jiri.svozilik@upol.cz} \affiliation{Palack\'{y} University,
RCPTM, Joint Laboratory of Optics, 17.listopadu 12, 771 46
Olomouc, Czech Republic}
\affiliation{ICFO-Institut de Ciencies
Fotoniques, Mediterranean Technology Park, 08860 Castelldefels,
Barcelona, Spain}

\author{Adam Vall\'{e}s}
 \affiliation{ICFO-Institut de
Ciencies Fotoniques, Mediterranean Technology Park, 08860 Castelldefels, 
Barcelona, Spain}

\author{Jan Pe\v{r}ina Jr.}
\affiliation{Palack\'{y} University, RCPTM, Joint Laboratory of
Optics, 17.listopadu 12, 771 46 Olomouc, Czech Republic}

\author{Juan P. Torres}
\affiliation{ICFO-Institut de Ciencies Fotoniques, Mediterranean
Technology Park, 08860 Castelldefels, Barcelona, Spain}
\affiliation{Department of Signal Theory and Communications, Universitat 
Politecnica de Catalunya, Campus Nord D3, 08034 Barcelona, Spain}

\date{\today}

\begin{abstract}
Coherence and correlations represent two related
properties of a compound system. The system can be, for instance,
the polarization of a photon, which forms part of a
polarization-entangled two-photon state, or the spatial shape of a
coherent beam, where each spatial mode bears different
polarizations. Whereas a local unitary transformation of the
system does not affect its coherence, global unitary transformations
modifying both the system and its surroundings can enhance its
coherence, transforming mutual correlations into coherence. The
question naturally arises of what is the best measure that
quantifies the correlations that can be turned into coherence, and
how much coherence can be extracted. We answer both questions, and
illustrate its application for some typical simple systems, with
the aim at illuminating the general concept of enhancing coherence
by modifying correlations.
\end{abstract}

\pacs{03.65.Ud,03.67.Mn,42.50.Dv,42.50.Ar}

\maketitle

\textit{Introduction.---}Coherence is one of the most
important concepts needed to describe the characteristics of a
stream of photons \cite{glauber1963,Mandel1995Optical}, 
where it allows us to characterize the interference capability of
interacting fields. However its use is far more general as
it plays a striking role in a whole range of physical, chemical,
and biological phenomena \cite{Chin2013Vibrational}. Measures of
coherence can be implemented using classical and quantum ideas,
which lead to the question of in which sense quantum coherence
might deviate from classical coherence phenomena
\cite{miller2012}, and to the evaluation of measures of coherence
\cite{levi2014,plenio2014,Streltsov2015Measuring}.

Commonly used coherence measures consider a physical
system as a whole, omitting its structure. The knowledge of \textit{the 
internal distribution of coherence} between
subsystems and their correlations becomes necessary for predicting
the evolution (migration) of coherence in the studied system. The
evolution of a twin beam from the near field into the far field
represents a typical example occurring in nature
\cite{Chan2007Transverse}. The creation of entangled states by
merging the initially separable incoherent and coherent states
serves as another example \cite{Streltsov2015Measuring}. Or, in
quantum computing the controlled-NOT gate entangles (disentangles)
two-qubit states \cite{OBrien2003Demostration,Nemoto2004Nearly},
at the expense (in favor) of coherence. Many quantum metrology and
communication applications benefit from correlations of entangled
photon pairs originating in spontaneous parametric down-conversion
\cite{Polyakov2009Quantum,Bouwmeester1997Experimental,Nielsen2010Quantum}.
Even separable states of photon pairs, i.e. states with suppressed
correlations, are very useful, e.g., in the heralded single photon
sources \cite{Mosley2008Heralded,Florez2015Correlation}. For all
of these, and many others, examples the understanding of common
evolution of coherence and correlations is crucial.

The Clauser-Horne-Shimony-Holt (CHSH) Bell's-like inequality
\cite{CHSH1969,bell1964,Horodecki1995Violating} has been usually considered
to quantify nonclassical correlations present between physically separated
photons that are entangled and so they can violate the bound set
by the inequality. However, correlations of a similar nature can
also exist when considering different degrees of freedom of a
single system \cite{gadway2009,valles2014}. The CHSH inequality
can also be violated when considering intrabeam correlations
between different degrees of freedom of intense beams, coherent or
not \cite{borges2010}. This, sometimes referred to as {\em nonquantum
entanglement}, or inseparability of degrees of freedom, has been
considered \cite{simon2010,eberly2011} as a tool to shed new light
into certain characteristics of classical fields, by applying
techniques usually restricted to a quantum scenario.

When the violation of the CHSH inequality between subsystems and the degree of
first-order coherence, which characterizes the internal coherence
of a physical subsystem \cite{glauber1963}, are combined together,
it is possible to define a measure that encompasses all coherences
and correlations in the system. This measure has been
experimentally examined by Kagalwala \textit{et al.}
\cite{Kagalwala2013Bells}. One fundamental problem of their
formulation is that it varies under global unitary
transformations. This means that, from this point of view, 
the amount of coherence in the system can be changed.

This behavior has several general consequences for any
partially coherent (mixed) state. First, the main
point is that the coherence of each subsystem can be increased by
means of a suitable unitary transformation affecting the whole
system. So the {\em hidden coherence} stored in the correlations
between two subsystems is made available. Second, for pure
states, the roles of the degree of entanglement between
subsystems, quantified by the concurrence
\cite{Hill1997Entanglement,wootters1998}, and the maximum violation of 
the CHSH inequality ($B_{\rm max}$) \cite{Horodecki1995Violating} are
interchangeable. However, this is not true for mixed states, where
the maximal violation can take place for states that are not
maximally entangled \cite{Zhou2013Entanglement}. This raises the
question of what is the appropriate measure to quantify {\em
hidden coherence} unveiled by global unitary transformations: the
degree of entanglement (concurrence) or $B_{\rm max}$.

In this Letter, we solve these two puzzles. First, given
a generally mixed state, or equivalently a partially coherent
light beam, {\em we determine what is the maximum and minimum
first-order coherence the subsystems can show under global unitary
transformations}. This will reveal how much hidden coherence is
present in the correlations between subsystems. Second,  we will
determine {\em if these maximal and minimal coherences are related
to states with the maximal (minimal) degree of entanglement, or
maximal or minimal violation of the CHSH inequality.} This will
solve the question of which of the two measures is the appropriate
one to quantify hidden coherence. Our main results are
expressed in two theorems valid for any mixed two-qubit quantum 
state, and their implication is illustrated by applying the
theorems to four well-known classes of quantum states.

We restrict our attention to coherence manipulations by a general
global unitary transformation. Experimentally, they can be
implemented by various logical gates
\cite{Nielsen2010Quantum,Knill2001Scheme,Lemr2015Experimental}.
The coherence limits can be also viewed as \textit{the maximal
coherence} that a logical gate can provide for a given state,
which is related to the entanglement power of a unitary operation
\cite{Guan2014Entangling}.

\textit{General considerations.---}Let us consider a $2
\times 2$ dimensional quantum state, $\hat{\rho}$, composed of
subsystems $A$ and $B$. The state $\hat{\rho}$ can be generally
written (spectral decomposition) as $\hat{\rho}=V\hat{E}V^{\dagger}$ 
\cite{Nielsen2010Quantum}, where $\hat{E}$ is a diagonal
matrix with eigenvalues that satisfy $\sum_i \lambda_i=1$ and $\lambda_1 \ge 
\lambda_2 \ge \lambda_3 \ge \lambda_4$. The matrix $V$ contains the 
corresponding eigenvectors. Each subsystem is characterized by the 
corresponding density matrix, $\hat{\rho}_A$ and $\hat{\rho}_B$. The degree of
first-order coherence of each subsystem is given 
$D_{A,B}=\sqrt{2\mathrm{Tr}[\hat{\rho}^2_{A,B}]-1}$ \cite{Mandel1995Optical}. 
We introduce here a measure of coherence for both subsystems when they are
considered independently $D^2=(D_A^2+D_B^2)/2$.  When both
subsystems are coherent, one has $D=1$, while only if both
subsystems show no coherence, $D=0$.

\textit{Minimum first-order coherence.---}There exists a
unitary transformation $U$ that when applied to $\hat{\rho}$
generates a new state
$\hat{\rho}^{\prime}=U\hat{\rho}U^{\dagger}$, so that the
coherence $D$ vanishes and the violation of the CHSH is maximized
with value \cite{Horodecki1995Violating,verstraete2002}
\begin{equation}
B_{\rm
max}=2\sqrt{2}\sqrt{\left(\lambda_1-\lambda_4\right)^2+\left(\lambda_2-\lambda_3\right)^2}.
\label{max_Bmax}
\end{equation}
The unitary transformation has the form $U=MV^{\dagger}$, where
\begin{equation}
M=\frac{1}{\sqrt{2}}\left(\begin{array}{cccc}
1 & 1 & 0 & 0 \\
0 & 0 & 1 & 1 \\
0 & 0 & 1 & -1\\
1 & -1 & 0 & 0
\end{array}\right).
\label{Eq.2}
\end{equation}
It is straightforward to show (see Supplemental Material) that
after the transformation $MV^{\dagger}$, $D_A=D_B=0$,
therefore $D=D_{\rm min}=0$. One can always achieve no coherence
for both subsystems. Therefore, the state with minimal coherence
is the state that provides maximal violation of the CHSH
inequality and it corresponds to the so-called Bell diagonal state
\cite{verstraete2002}.

The degree of entanglement (concurrence) of Bell diagonal states
is $C_{BD}=\max \left\{0,2\lambda_1-1
\right\}$ \cite{verstraete2002}. The maximum concurrence that can be achieved 
by a unitary operation applied on $\hat{\rho}$ is $C_{\rm
max}=\max\left\{0,\lambda_1-\lambda_3-2\sqrt{\lambda_2
\lambda_4}\right\}$ \cite{verstraete2001}. As we will see in example I, $C_{BD} 
\le C_{\rm max}$ can happen for mixed states,
which highlights the preference for using $B_{\rm max}$ over the
concurrence for quantifying the coherence available for each
subsystem.

\textit{Maximum first-order coherence.---}There exists a
unitary transformation $U$ that when applied to an arbitrary state
$\hat{\rho}$ generates a new state
$\hat{\rho}^{\prime}=U\hat{\rho}U^{\dagger}$ that maximizes the coherence
$D$ with value
\begin{equation}
D^2_{\rm
max}=\left(\lambda_1-\lambda_4\right)^2+\left(\lambda_2-\lambda_3\right)^2
\label{max_D}
\end{equation}
and yields a violation of CHSH that is minimal, with value
\begin{equation}
B_{\rm max}=2\left|\lambda_1-\lambda_2-
\lambda_3+\lambda_4\right|.
\label{min_Bmax}
\end{equation}
The unitary transformation $U$ has the form $U=V^{\dagger}$.

The resulting state is a diagonal separable state, as it is shown in the
Supplemental Material.

$D_{\rm max}$ can be called the {\em degree of available
coherence}, since it represents the maximum first-order coherence
that can be unveiled under a global unitary transformation.
As we will show in example I below,
correlations can be a source of coherence for a subsystem even
when the CHSH inequality is not violated, i.e., $B_{\rm max} \le
2$, and therefore the state is not entangled. Importantly, $D_{\rm
max}$ is associated to a state with the minimum violation of the
CHSH inequality, highlighting again the outstanding role of
$B_{\rm max}$ over concurrence when considering the maximum and
minimum values of the degree of coherence available.

\begin{figure}[t!]
\includegraphics[width=8.5cm]{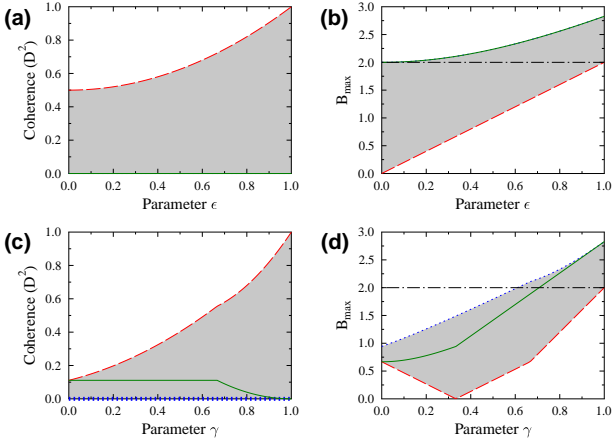}
 \caption{Coherence ($D^2$) and maximal violation of the CHSH inequality
 ($B_{\rm max}$) for (a) and (b): example I ($\hat{\rho}_{\rm MNMS}$), and (c) 
 and (d):  example II ($\hat{\rho}_{\rm MEMS}$). Green lines depict the values 
 of the original state, prior to any unitary transformation.
 The maximal coherence and minimal violation of the CHSH inequality
 are marked by dashed-red lines, and the minimal coherence and maximal 
 violation  of the CHSH inequality are marked by dotted-blue lines.
 The black dashed-dotted line represents the classical correlation limit 
 $B_{\rm max}=2$. Grey areas correspond to all admissible values
 achievable by all unitary operations.}
 \label{Fig1}
\end{figure}

We will now consider four examples where we apply the results
mentioned above.

\begin{figure}[t!]
\includegraphics[width=8.5cm]{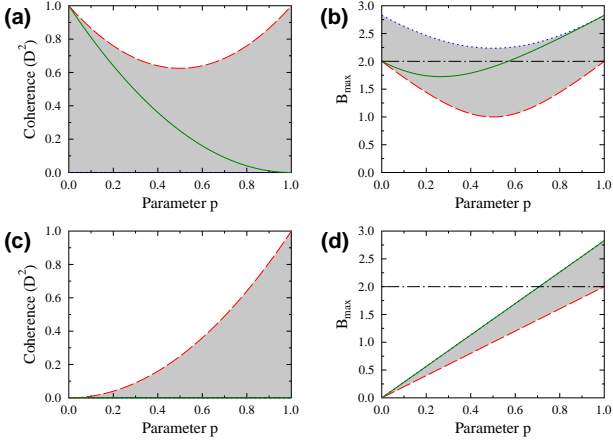}
 \caption{Coherence ($D^2$) and maximal violation of the CHSH inequality
 ($B_{\rm max}$) for (a) and (b): example III ($\hat{\rho}_{\rm EXC}$); and (c) 
 and (d): example IV ($\hat{\rho}_{\rm W}$).
 Green lines depict the values of the original state, prior to any unitary
 transformation. The maximal coherence and minimal violation of the CHSH 
 inequality are marked by  dashed-red lines, and the minimal coherence and 
 maximal violation of the CHSH inequality are marked by dotted-blue lines.
 The black dashed-dotted line represents the classical correlation limit
 $B_{\rm max}=2$. Grey areas correspond to all admissible values achievable by all unitary
 operations.}
 \label{Fig2}
\end{figure}

\textit{Example I: Maximally nonlocal mixed state (MNMS).---}
In a nonlinear process designed to generate entanglement in polarization
\cite{kwiat1995,torres2011}, the state generated at the output of
the nonlinear crystal can be generally written in the
computational basis
$\{|00\rangle,|01\rangle,|10\rangle,|11\rangle\}$ as
\cite{Batle2011Nonlocality,Zhou2013Entanglement}
\begin{equation}\hat{\rho}_{\rm MNMS}=\left(%
\begin{array}{cccc}
  1/2 & 0 & 0 & \epsilon/2 \\
  0 & 0 & 0 & 0 \\
  0 & 0 & 0 & 0 \\
  \epsilon/2 & 0 & 0 & 1/2 \\
\end{array}%
\right)\;\;\mathrm{where}\;\;\epsilon\in\langle 0,1\rangle.
\label{Eq:MNM}
\end{equation}
The purity of the state is $P={\rm
Tr}[\hat{\rho}_{\rm MNMS}^2]=(1+\epsilon^2)/2$. The spectral
representation of this state writes
$\hat{\rho}_{\rm 
MNMS}=1/2\,(1+\epsilon)|\Phi^+\rangle\langle\Phi^+|+1/2\,(1-\epsilon)|\Phi^-\rangle\langle\Phi^-|$
This state is a Bell diagonal state, so it produces a maximal
violation of the CHSH inequality. For each value of $\epsilon$,
the state $\hat{\rho}_{\rm MNMS}$ can be transformed using unitary
operations to a new state $\hat{\rho}'_{\rm MNMS}$ with new values of
$D^2$ [see Fig. \ref{Fig1}(a)] and $B_{\rm max}$ [see Fig.
\ref{Fig1}(b)]. The grey areas in the figures show all possible
values of $D^2$ and $B_{\rm max}$. In all cases presented here,
and shown in Figs. \ref{Fig1}--\ref{Fig2}, we performed extensive
numerical simulations \cite{Jarlskog2005Recursive} generating
$10^6$ randomly generated unitary operations for each value of
parameters, to check all of our predictions.

All of these values lie in intervals limited by states with
minimal and maximal coherence. The state already yields minimal
coherence ($D_A=D_B=0$) and maximal violation of the CHSH
inequality, as given by Eq. (\ref{max_Bmax}) [dotted-blue lines in
Figs. \ref{Fig1}(a) and \ref{Fig1}(b)]
\begin{equation}
D_{A}=D_B=0 , \hspace{0.5cm} B_{\rm max}=2\sqrt{1+\epsilon^2}.
\end{equation}
The case of maximal coherence and minimal violation of the CHSH
inequality is given by Eqs. (\ref{max_D}) and (\ref{min_Bmax})
[dashed-red lines in Figs. \ref{Fig1}(a) and \ref{Fig1}(b)]
\begin{equation}
D_{\rm max}^2=\frac{1+\epsilon^2}{2}, \hspace{0.5cm} B_{\rm
max}=2|\epsilon|.
\end{equation}
The degree of entanglement of the quantum state with minimum
first-order coherence ($D_A=D_B=0$), which corresponds to the
maximal violation of the CHSH inequality, is $C_{BD}=\epsilon$.
However, the maximum entanglement that can be achieved with a
unitary operation is $C_{\rm max}=(1+\epsilon)/2$. Therefore
$C_{BD} < C_{\rm max}$. This shows the relevant role $B_{\rm max}$
over the concurrence. The state which achieves minimal first-order
coherence for a subsystem is also the state that maximally
violates the CHSH inequality, but not the state that achieves
maximum entanglement.

\textit{Example II: Maximally entangled mixed state (MEMS)---}This state is 
defined as \cite{Munro2001Maximizing,wei2003}
\begin{equation}
\hat{\rho}_{MEMS}=\left\{\begin{array}{cc}
\left(\begin{array}{cccc}
1/3 & 0 & 0 & \gamma/2\\
0 & 1/3 & 0 & 0 \\
0 & 0 & 0 & 0 \\
\gamma/2 & 0 & 0 & 1/3
\end{array}\right) & \mathrm{for}\; 0\leq \gamma \leq \frac{2}{3}\\
&\\
\left(\begin{array}{cccc}
\gamma/2 & 0 & 0 & \gamma/2\\
0 & 1-\gamma & 0 & 0 \\
0 & 0 & 0 & 0 \\
\gamma/2 & 0 & 0 & \gamma/2
\end{array}\right) & \mathrm{for}\; \frac{2}{3}\leq \gamma\leq 1
\end{array}\right..
\label{Eq:MEMS}
\end{equation}

\noindent It maximizes the value of the concurrence for a given value of the
purity. We have chosen the phases to be zero for the sake of
simplicity. The purity is equal to
$P=\frac{1}{3}+\frac{\gamma^2}{2}$ for
$0\leq\gamma\leq\frac{2}{3}$ and
$P=\gamma^2+\left(1-\gamma\right)^2$ for
$\frac{2}{3}\leq\gamma\leq1$. When the state is transformed to the
new state using unitary operations [see Figs. \ref{Fig1}(c) and \ref{Fig1}(d)],
we find that for $0\leq \gamma \leq \frac{2}{3}$ the minimal
coherence and maximal violation of the CHSH are [dotted-blue lines
in Figs. \ref{Fig1}(c) and \ref{Fig1}(d)]
\begin{equation}
D_A=D_B=0, \hspace{0.5cm} B_{\rm
max}=2\sqrt{2}\sqrt{\frac{\gamma^2}{4}+\left(\frac{1}{3}+\frac{\gamma}{2}\right)^2}
\end{equation}
and the maximal coherence and minimal violation of the CHSH are
[dashed-red lines in Figs. \ref{Fig1}(c) and \ref{Fig1}(d)]
\begin{equation}
D_{\rm
max}^2=\frac{\gamma^2}{4}+\left(\frac{1}{3}+\frac{\gamma}{2}\right)^2,
\hspace{0.4cm} B_{\rm max}=2\left|\gamma-\frac{1}{3}\right|.
\end{equation}
For $\frac{2}{3}\leq \gamma \leq 1$, these limits are [dotted-blue
and dashed-red lines in Figs. \ref{Fig1}(c) and \ref{Fig1}(d)]
\begin{equation}
D_A=D_B=0, \hspace{0.5cm} B_{\rm
max}=2\sqrt{2}\sqrt{\gamma^2+\left(1-\gamma\right)^2}
\end{equation}
and
\begin{equation}
D_{\rm max}^2=\gamma^2+\left(1-\gamma \right)^2, \hspace{0.4cm}
B_{\rm max}=2|2\gamma-1|.
\end{equation}
The green lines in Figs. \ref{Fig1}(c) and \ref{Fig1}(d) show the actual
value of $D^2$ and $B_{\rm max}$, prior to the application of any
unitary transformation.

\textit{Example III: State considered in 
\cite{Kagalwala2013Bells}.---}Kagalwala \textit{et al.} investigated
(example C) a state whose density matrix writes
\begin{equation}
\hat{\rho}_{EXC}\left(p\right)=\frac{1}{2}\left(\begin{array}{cccc}
1-p & 0 & 1-p & 0\\
0 & p & ip & 0\\
1-p & -ip & 1 & 0\\
0 & 0 & 0 & 0
\end{array}\right) \; \mathrm{where}\;  p\in\langle0,1\rangle.
\label{Eq:EXC}
\end{equation}
The purity of this state is $ P=1-\frac{3}{2}p+\frac{3}{2} p^2$.
In Figs. \ref{Fig2}(a) and \ref{Fig2}(b) all possible values of
$D^2$ and $B_{\rm max}$ are shown for this particular case. The boundaries
of the grey areas are formed by the states with minimal coherence
and maximal violation of the CHSH inequality \cite{comment2}
\begin{equation}
D_1=D_2=0, \hspace{0.5cm} B_{\rm
max}=2\sqrt{2}\,\sqrt{1-\frac{3}{2}p+\frac{3}{2}p^2}
\end{equation}
and the maximal coherence and correspondingly minimal violation of the
CHSH inequality
\begin{equation}
D_{\rm max}^2=1-\frac{3}{2}p+\frac{3}{2}p^2, \hspace{0.3cm} B_{\rm
max}=2\sqrt{1-3p+3p^2}.
\end{equation}

\textit{Example IV: Werner state.---}As a final example we
consider the Werner state \cite{Werner1989Quantum}, which is
defined as
\begin{equation}
\hat{\rho}_{\rm W}\left(p\right)=\frac{1}{4}\left(\begin{array}{cccc}
1+p & 0 & 0 & 2p\\
0 & 1-p & 0 & 0\\
0 & 0 & 1-p & 0\\
2p & 0 & 0 & 1+p
\end{array}\right)\;\mathrm{where}\; p \in \langle0,1\rangle.
\label{Eq:WER}
\end{equation}
The purity is $P=(1+3p^2)/4$. When this state is transformed,
$D^2$ and $B_{\rm max}$ can attain any value inside the grey areas
in Figs. \ref{Fig2}(c) and \ref{Fig2}(d). For these plots, the limits are
\begin{equation}
D_1=D_2=0, \hspace{0.5cm} B_{\rm max}=2\sqrt{2}p
\end{equation}
for minimal coherence and maximal violation of the CHSH inequality
and
\begin{equation}
D_{\rm max}^2=p^2, \hspace{0.4cm} B_{\rm max}=2 p,
\end{equation}
for maximal coherence and minimal violation of the CHSH inequality

\textit{The relationship between coherence and
correlations.---}For a given quantum state, the relationship between the
degree of coherence of each subsystem and the correlations between
subsystems is quantified by the measure $S_{A,B}=D_{A,B}^2/2+\left(B_{\rm max}/2\sqrt{2}\right)^2$
called \textit{accessible coherence} in the subsystem $A$, $B$ 
\cite{Kagalwala2013Bells}.
Especially, for a pure state the statement
\begin{equation}
\frac{D_{A,B}^2}{2}+\left(\frac{B_{\rm max}}{2\sqrt{2}}
\right)^2=1 \label{Eq:t1}
\end{equation}
is valid. Any increase (or decrease) of the degree of coherence is
compensated by a corresponding change of $B_{\rm max}$. This
relationship is no longer true for mixed states as shown in the
Supplemental Material.

What is then, for all states, the appropriate equation that
relates first-order coherence and correlations? For a generally
mixed state ($\mathrm{Tr} \hat{\rho}^2 \le 1$), one can derive
\cite{Horodecki1995Violating}
\begin{equation}
\frac{D_A^2+D_B^2}{4}+\mathcal{T}=\mathrm{Tr} \hat{\rho}^2,
\label{Eq:t2}
\end{equation}
where $\mathcal{T}=1/4\,(1+\sum_{i,j=1}^3 t_{ij}^2)$,
$t_{ij}=\mathrm{Tr}\left[\hat{\rho}\hat{\sigma}_i\otimes\hat{\sigma}_j\right]$,
and $\sigma_{i,j}$ ($i,j=1,2,3$) are Pauli matrices. The values of
$t_{ij}$ can only be obtained by making coincidence measurements
between the subsystems, therefore measuring the nature of its
correlations. In general
\begin{equation}
 \frac{(\lambda_1+\lambda_4)^2+(\lambda_2+\lambda_3)^2}{2} \leq
 \mathcal{T}
 \leq
 \mathrm{Tr} \hat{\rho}^2.
\end{equation}
For a pure state, $D_A=D_B$ and $\mathcal{T}=(B_{\rm
max}/2\sqrt{2})^2$, so one obtains
Eq. (\ref{Eq:t1}). For maximally entangled states, $B_{\rm
max}=2\sqrt{2}$, so $\mathcal{T}=1$ achieves its maximum value,
while for separable pure states, $B_{\rm max}=2$ and
$\mathcal{T}=1/2$.

\textit{Conclusions.---}We have solved several puzzles
about the relationship between coherence and certain measures of
correlations present between subsystems, as it is the case of the
CHSH inequality. For the case of two correlated two-dimensional subsystems,
we have obtained simple expressions that quantify the amount of
first-order coherence that can be obtained in each subsystem ({\em
hidden coherence}) by modifying correlations between the
subsystems. We have shown that the relevant parameter to quantify
the maximum hidden coherence is the degree of violation of the
CHSH inequality, not the degree of entanglement between
subsystems. Although we have considered here only a few systems as
examples, their analysis, based on suitably defined quantities,
illuminates the general concept of extracting coherence from
manipulating the correlations between subsystems.

\begin{acknowledgments}
We thank A. Miranowicz and M. Oszmaniec for discussions. This work
was supported by Severo Ochoa (Government of Spain) and Fundacio
Privada Cellex Barcelona. J.S. and J.P. acknowledge the project 
CZ.1.07/2.3.00/30.0004 of the Ministry of Education, Youth and Sports of the 
Czech Republic, and the project 15-08971S of the Czech Science
 Foundation.
\end{acknowledgments}

\end{document}

\title{Revealing Hidden Coherence in Partially Coherent Light\\
Supplemental Material}

\author{Ji\v{r}\'{i} Svozil\'{i}k}
\email{jiri.svozilik@upol.cz} \affiliation{Palack\'{y} University,
	RCPTM, Joint Laboratory of Optics, 17.listopadu 12, 771 46
	Olomouc, Czech Republic}
\affiliation{ICFO-Institut de Ciencies
	Fotoniques, Mediterranean Technology Park, 08860 Castelldefels,
	Barcelona, Spain}

\author{Adam Vall\'{e}s}
\affiliation{ICFO-Institut de
	Ciencies Fotoniques, Mediterranean Technology Park, 08860 Castelldefels, 
	Barcelona, Spain}

\author{Jan Pe\v{r}ina Jr.}
\affiliation{Palack\'{y} University, RCPTM, Joint Laboratory of
	Optics, 17.listopadu 12, 771 46 Olomouc, Czech Republic}

\author{Juan P. Torres}
\affiliation{ICFO-Institut de Ciencies Fotoniques, Mediterranean
	Technology Park, 08860 Castelldefels, Barcelona, Spain}
\affiliation{Department of Signal Theory and Communications, Universitat 
	Politecnica de Catalunya, Campus Nord D3, 08034 Barcelona, Spain}

\maketitle

Let $\hat{\rho}$ be a general $4$ dimensional complex Hermitian
matrix, that could represent a $2 \times 2$-dimensional mixed
quantum state, or a partially coherent beam describing coherent or
incoherent superpositions of four modes. In general, $\hat{\rho}$
can be always written as ({\em spectral decomposition})
\begin{equation}  \hat{\rho}=V\hat{E}V^{\dagger}=\lambda_1
|a\rangle\langle a|+ \lambda_2 |b\rangle\langle b|+\lambda_3
|c\rangle\langle c|+\lambda_4 |d\rangle\langle d|,
\end{equation}
where the diagonal matrix $\hat{E}$ contains eigenvalues
$\lambda_i$ and the matrix V consists of the corresponding
eigenvectors $\left\{ |a\rangle, |b\rangle, |c\rangle, |d\rangle
\right\}$ forming an orthonormal basis. We assume that $\lambda_1
\geq \lambda_2 \geq \lambda_3 \geq \lambda_4$ and $\sum_i
\lambda_i=1$. When a unitary transformation U is applied to the
state $ \hat{\rho} $, a new state $\hat{\rho}^{\prime}=U
\hat{\rho} U^{\dagger}$ is obtained,
\begin{equation}
\hat{\rho}^{\prime}=\lambda_1 |a^{\prime}\rangle\langle
a^{\prime}|+ \lambda_2 |b^{\prime}\rangle\langle
b^{\prime}|+\lambda_3 |c^{\prime}\rangle\langle
c^{\prime}|+\lambda_4 |d^{\prime}\rangle\langle d^{\prime}|.
\end{equation}
Notice that all states connected by the means of unitary
transformations share the same eigenvalues, i.e., the eigenvalues
$\lambda_i$ are invariant under the unitary transformations.
However, the eigenvectors change.

\section{Minimal first-order coherence}
\noindent\textbf{Theorem:}

There exists a unitary transformation $U$ that when applied to
$\hat{\rho}$ generates a new state
$\hat{\rho}^{\prime}=U\hat{\rho}U^{\dagger}$ so that the coherence
$D$ vanishes and the violation of the CHSH is maximized with value
\cite{Horodecki1995Violating,verstraete2002}
\begin{equation}
B_{\rm
max}=2\sqrt{2}\sqrt{\left(\lambda_1-\lambda_4\right)^2+\left(\lambda_2-\lambda_3\right)^2}.
\label{Eq.1}
\end{equation}
The unitary transformation has the form $U=MV^{\dagger}$, where
\begin{equation}
M=\frac{1}{\sqrt{2}}\left(\begin{array}{cccc}
1 & 1 & 0 & 0 \\
0 & 0 & 1 & 1 \\
0 & 0 & 1 & -1\\
1 & -1 & 0 & 0
\end{array}\right).
\label{Eq.2}
\end{equation}

\noindent\textbf{Proof:}

First, we need to transform $\hat{\rho}$ to a diagonal form in the
computational basis $\{|0\rangle_A |0\rangle_B, |0\rangle_A
|1\rangle_B, |1\rangle_A |0\rangle_B, |1\rangle_A |1\rangle_B \}$.
This is done with the help of the matrix $V$ that contains the
eigenvectors of $\hat{\rho}$, so that
\begin{equation}
\hat{\rho} \rightarrow \hat{E}=V^{\dagger}
\hat{\rho}V.
\end{equation}
From \cite{verstraete2002}, it can be shown that the violation of
the CHSH inequality is maximized for a Bell diagonal state of the
form
\begin{eqnarray}
& & \hat{\rho}=\lambda_1|\Phi^{+}\rangle\langle\Phi^{+}|+
\lambda_2|\Phi^{-}\rangle\langle\Phi^{-}| \nonumber
\\
& & +\lambda_3|\Psi^{+}\rangle\langle\Psi^{+}|
+\lambda_4|\Psi^{-}\rangle\langle\Psi^{-}|, \label{bell1}
\end{eqnarray}
where $|\Phi^{\pm}\rangle$ and $|\Psi^{\pm}\rangle$ are the
maximally entangled Bell states. Any unitary transformation
applied on the state given by Eq. (\ref{bell1}) cannot increase
the degree of violation of the inequality.

A general Bell diagonal state in the computational basis writes
as
\begin{equation}
\hat{\rho}_{\rm Bell}=\frac{1}{2}\left(\begin{array}{cccc}
\lambda_1+\lambda_2 & 0 & 0 & \lambda_1-\lambda_2\\
0 & \lambda_3+\lambda_4 &\lambda_3-\lambda_4 & 0\\
0 & \lambda_3-\lambda_4 & \lambda_3+\lambda_4 & 0\\
\lambda_1-\lambda_2 & 0 & 0 & \lambda_1+\lambda_2\\
\end{array}\right).
\label{Eq.3}
\end{equation}
The matrix $M$ performs the transformation
\cite{Thirring2011Entanglement}
\begin{equation}
\left\{ |00\rangle,|01\rangle,|10\rangle,|11\rangle \right\}
\Longrightarrow \left\{ |\Phi^{+}\rangle, |\Phi^{-}\rangle,
|\Psi^{+}\rangle, |\Psi^{-}\rangle \right\}
\end{equation}
that can be easily demonstrated by direct inspection. After
combining both transformations, now we have
\begin{equation}
\hat{\rho} \Longrightarrow \hat{\rho}^{\prime}=M \hat{E}
M^{\dagger}=M V^{\dagger} \hat{\rho} V M^{\dagger}.
\end{equation}
The unitary transformation $U=M V^{\dagger}$ generates the state given in
Eq. (\ref{bell1}). From here, one can use the Horodecki's approach
\cite{Horodecki1995Violating} to get Eq. (\ref{Eq.1}) \cite{verstraete2002}.

The coherence ($D$) for the state of the form given in
Eq. (\ref{Eq.3}) is
\begin{equation}
D_A^2=2 \mathrm{Tr} \left[\frac{1}{4}\left(\begin{array}{cc}
\sum_{i=1}^4 \lambda_i & 0 \\
0 & \sum_{i=1}^4 \lambda_i
\end{array}\right)^2\right]-1=0.
\end{equation}
Similarly, we obtain $D_B^2=0$. That leads to
$D^2=(D_A^2+D_B^2)/2=0$.

By means of a unitary transformation, we can generate a new state
where both subsystems show no coherence $D=0$. It corresponds to
the state that shows the maximal violation of the CHSH inequality
achievable for states connected through the unitary
transformations.

\section{Maximal first-order coherence}
\noindent\textbf{Theorem:}

There exists a unitary transformation $U$ that when applied to an
arbitrary state $\hat{\rho}$ generates a new state
$\hat{\rho}^{\prime}=U\hat{\rho}U^{\dagger}$ that maximizes the
coherence $D$ with value
\begin{equation}
D^2_{\rm
max}=\left(\lambda_1-\lambda_4\right)^2+\left(\lambda_2-\lambda_3\right)^2
\label{max_D}
\end{equation}
and yields a violation of CHSH that is minimal, with value
\begin{equation}
B_{\rm max}=2\left|\lambda_1-\lambda_2-\lambda_3+\lambda_4\right|.
\label{min_Bmax}
\end{equation}
The unitary transformation $U$ has the form $U=V^{\dagger}$.

\noindent\textbf{Proof:}

The unitary transformation $U=V^{\dagger}$ transforms an arbitrary state
$\hat\rho $ into the state $\hat{E}$ that is diagonal in the
computational basis $\{|0\rangle_A |0\rangle_B,|0\rangle_A
|1\rangle_B, |1\rangle_A |0\rangle_B, |1\rangle_A |1\rangle_B,\}$,
so it performs the transformation
\begin{equation}
\left\{ |a\rangle, |b\rangle, |c\rangle, |d\rangle \right\}
\Longrightarrow \left\{ |0\rangle_A |0\rangle_B, |0\rangle_A
|1\rangle_B, |1\rangle_A |0\rangle_B, |1\rangle_A |1\rangle_B
\right\}.
\end{equation}
Therefore
\begin{equation}
\hat{\rho} \Longrightarrow \hat{\rho}^{\prime}=V^{\dagger} \hat{\rho}
V.
\end{equation}
One can see by performing extensive numerical simulations that the
degree of coherence $D$ cannot be increased by applying additional
unitary transformations $W$ on $\hat{E}$.

Moreover, when considering the Jarlskog recursive parametrization
\cite{Jarlskog2005Recursive} of an arbitrary unitary
transformation $W(\vec{\alpha})$ with parameter $\vec{\alpha}$, we
can demonstrate that the function that gives the degree of
coherence after the unitary transformation, $D[W(\vec{\alpha})
\hat{E}W^{\dagger}(\vec{\alpha})]$ has a maximum for
$\vec{\alpha}=0$, which corresponds to the identity
transformation, since
\begin{equation}
\left.\frac{\partial
D\left[W(\vec{\alpha})\hat{E}W^{\dagger}(\vec{\alpha})\right]}{\partial\vec{\alpha}}\right|_{\vec{\alpha}=\vec{0}}=0.
\end{equation}
Direct calculation of the Hessian matrix of second derivatives
confirms that the state $\hat{E}$ has the maximal degree of
coherence $D$. In this proof, alternating signs of the
determinants of leading sub-matrices with the increasing rank have
been obtained.

The degree of coherence $D$ of the state $\hat E$ with diagonal
elements $\lambda_i$ is easily obtained to be
\begin{equation}
 D^2_{\rm
 max}=\frac{D_A^2+D_B^2}{2}=\left(\lambda_1-\lambda_4\right)^2+\left(\lambda_2-\lambda_3\right)^2.
\label{equation17}
\end{equation}

For the state $\hat E$, the only non-vanishing element of the
matrix $T_{\rho}$ is $t_{33}$, that reads
\begin{equation}
T_{\hat{E}}=\left(\begin{array}{ccc}
0 & 0 & 0 \\
0 & 0 & 0 \\
0 & 0 & \lambda_1-\lambda_2-\lambda_3+\lambda_4
\end{array}
\right).
\end{equation}
The value of $B_{\rm max}$ is $B_{\rm max}=2\sqrt{\mu}$, where
$\mu=(\lambda_1-\lambda_2-\lambda_3+\lambda_4)^2$ is the only
non-zero eigenvalue of $T_{\hat{E}}^{T}T_{\hat{E}}$. In this case
\begin{equation}
B_{\rm max}=2|\lambda_1-\lambda_2-\lambda_3+\lambda_4|.
\label{equation19}
\end{equation}$ $

\begin{figure}[t!]
\includegraphics[width=9.2cm]{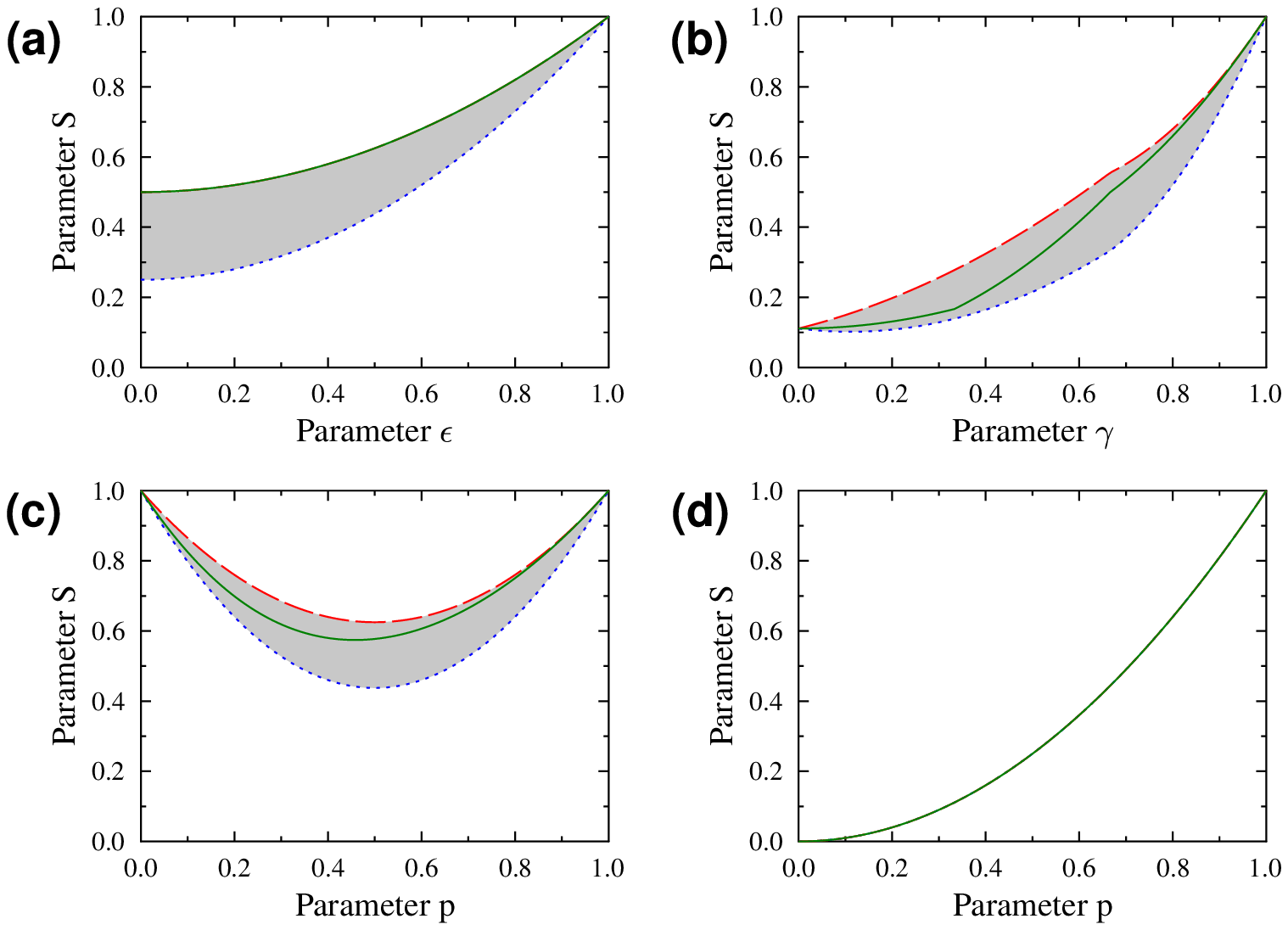}
\caption{Maximum and minimum values of $S$ for examples I-IV. (a)
MNMS state, (b) MEMS state, (c) example C in
\cite{Kagalwala2013Bells} and (d) Werner state. The green line
depicts the original value, prior to any unitary transformation.
The dashed-red line depicts $S_{\rm max}$, and the dotted-blued
line depicts $S_{\rm min}$. Grey areas mark all possible values of
$S$. } \label{Fig:3}
\end{figure}

\section{General invariant involving coherence}
One can be tempted to look for an expression similar to the one
defined in \cite{Kagalwala2013Bells} and define a parameter $S$ as
\begin{equation}
S=\frac{D_A^2+D_B^2}{4}+\left(\frac{B_{\rm max}}{2\sqrt{2}}
\right)^2. \label{Eq:t10}
\end{equation}
For certain states, it can be found that this parameter is indeed
constant under unitary transformations. This is the case of the
Werner state (example IV in the main text), as it is demonstrated
below and can be observed in Fig. \ref{Fig:3}(d). However, in
general, this is not the case. Examples I-III of the main text
correspond to this situation, as it can be seen in Figs.
\ref{Fig:3}(a)-(c), that show all possible values of $S$ (grey
areas) obtained by unitary transformations.

It can be easily shown using Eqs.(1)-(4) of the main text that all
values of $S$ are between upper and lower boundaries: $S_{\rm
max}= P-2(\lambda_1 \lambda_4+\lambda_2 \lambda_3)$ and $S_{\rm
min}= P-(\lambda_1 +\lambda_4)(\lambda_2+\lambda_3)$, where $P$
stands for the purity of the state. $S_{\rm max}$ corresponds to
the maximal violation of the CHSH inequality and the minimal first-order
coherence, whereas $S_{\rm min}$ corresponds to the minimal violation
of the CHSH inequality and the maximal first-order coherence.

For a general mixed state ($\mathrm{Tr} \hat{\rho}^2 \le 1$), one
can derive \cite{Horodecki1995Violating}
\begin{equation}
    \frac{D_A^2+D_B^2}{4}+\mathcal{T}=\mathrm{Tr} \hat{\rho}^2,
    \label{Eq:t2}
\end{equation}
where $\mathcal{T}=1/4\,(1+\sum_{i,j=1}^3 t_{ij}^2)$,
$t_{ij}=\mathrm{Tr}\left[\hat{\rho}\hat{\sigma}_i\otimes\hat{\sigma}_j\right]$
and $\sigma_{i,j}$ ($i,j=1,2,3$) are the Pauli matrices. The
values of $t_{ij}$ can be only obtained by making coincidence
measurements between the subsystems, therefore measuring the
nature of their correlations. Any increase/decrease of the degree of
coherence is accompanied by a corresponding decrease/increase of
$\mathcal{T}$. For a pure state, $D_A=D_B$ and
$\mathcal{T}=(B_{\rm
    max}/2\sqrt{2})^2$ (see below), so one obtains
\begin{equation}
\frac{D_{A,B}^2}{2}+\left(\frac{B_{\rm max}}{2\sqrt{2}} \right)^2=1.
\label{Eq:t1}
\end{equation}

\section{Pure states: Derivation of Eq. (\ref{Eq:t1}) }
Any pure state can be written as a Schmidt decomposition that
reads \cite{peres}
\begin{equation}
|\Psi\rangle=\kappa_1 |x_1\rangle_A
|y_1\rangle_B+\kappa_2|x_2\rangle_A |y_2\rangle_B,
\label{Eq:31}
\end{equation}
where $\kappa_1^2+\kappa_2^2=1$, $\{ |x_1\rangle_A ,|x_2\rangle_A
\}$ is an orthonormal basis in subsystem $A$ and $\{
|y_1\rangle_B, |y_2\rangle_B \}$ is an orthonormal basis in
subsystem $B$. For the sake of simplicity we assume $\kappa_{1,2}$
to be real. Following \cite{Horodecki1996Violating}, one obtains
that
\begin{eqnarray}
& & \sum_{i,j=1}^3 t_{ij}^2=1+8\kappa_1^2\kappa_2^2, \\
& & B_{\rm max}=2\sqrt{1+4\kappa_1^2\kappa_2^2}.
\end{eqnarray}
Therefore
\begin{eqnarray}
& & \mathcal{T}=\frac{1+\sum_{i,j=1}^3
t_{ij}^2}{4}=\frac{1+4\kappa_1^2\kappa_2^2}{2}
\nonumber \\
& &  =\frac{1}{2}\left[ 1+\left(\frac{B_{\rm max}^2}{4}-1
\right)\right]=\left(\frac{B_{\rm max}}{2\sqrt{2}}\right)^2
\label{sup1}
\end{eqnarray}
Substitution of Eq. (\ref{sup1}) into Eq. (\ref{Eq:t2})
yields straightforwardly Eq. (\ref{Eq:t1}).

\section{Werner state: Invariance of the parameter $S$ under global unitary transformations}
The Werner state (example IV of the main paper) can be written as
\cite{hiroshima2000}
\begin{equation}
\hat{\rho}_{W}=\frac{1-p}{4} I_4+p|\Phi^{+}\rangle\langle \Phi^+|,
\end{equation}
where $|\Phi^{+}\rangle=1/\sqrt{2}\,(|0\rangle_A
|0\rangle_B+|1\rangle_A |1\rangle_B)$ is the maximally entangled
Bell state. The Werner state can be generalized considering the
state $|\Psi\rangle$ given by Eq. (\ref{Eq:31}) instead of
$|\Phi^{+}\rangle$. The spectral decomposition of the generalized
state can be written as
\begin{eqnarray}
& & \hat{\rho}_{W}=\frac{1+3p}{4} |\Psi\rangle\langle\Psi|+
\frac{1-p}{4}|\Psi^{\perp}\rangle\langle\Psi^{\perp}|\\
& & + \frac{1-p}{4}\left(|x_1\rangle_A|y_2\rangle_B\langle x_1|_A
\langle y_2|_B +|x_2\rangle_A |y_1\rangle_B\langle x_2|_A \langle
y_1|_B\right), \nonumber
\end{eqnarray}
where
$|\Psi^{\perp}\rangle=-\kappa_2|x_1\rangle_A|y_1\rangle_B+\kappa_1|x_2\rangle_A|y_2\rangle_B$.
The corresponding matrix $T_{\rho}$ writes
\begin{equation}
T_{\rho}=\left(\begin{array}{ccc}
2p\kappa_1 \kappa_2 & 0 & 0 \\
0 & -2p\kappa_1 \kappa_2 & 0 \\
0 & 0 & p
\end{array}
\right).
\end{equation}
The maximal violation of the CHSH inequality writes
\begin{equation}
B_{\rm max}=2p\sqrt{1+4 \kappa_1^2 \kappa_2^2}
\label{B_max_werner}
\end{equation}
and
\begin{equation}
\mathcal{T}=\frac{1+p^2+8p^2 \kappa_1^2 \kappa_2^2}{4} \label{T}.
\end{equation}
Substituting  Eq. (\ref{T}) into Eq. (\ref{Eq:t2}), and
making use of Eq. (\ref{B_max_werner}), one obtains that
\begin{equation}
\frac{D^2}{2}+\left(\frac{B_{\rm max}}{2\sqrt{2}}\right)^2=p^2.
\end{equation}
For $p=1$ (pure state) we recover Eq. (\ref{Eq:t1}).